# A Model for Privacy-enhanced Federated Identity Management

Rainer Hörbe, Identinetics GmbH

19 January 2014

**Abstract**

Identity federations operating in a business or consumer context need to prevent the collection of user data across trust service providers for legal and business case reasons. Legal reasons are given by data protection legislation such as [1]. Other reasons include business owners becoming increasingly aware of confidentiality risks that go beyond traditional information security, e.g., the numbers of authentications to an EDI service might provide insights into the volume of invoices, from which one could derive insider information. This paper proposes extended technical controls supporting three privacy requirements:

a) Limited Linkability: Two service providers cannot link data related to a user without the help of a third party, using neither an identifier nor other identifying attributes like email addresses or payment data;

b) Limited Observability: An identity provider cannot trace which services a user is using without the help of a third party;

c) Non-Disclosure: Attributes provided to the service provider by an attribute provider are not disclosed to the identity provider or an intermediate service broker.

Using a hub-and-spoke federation style following the privacy-by-design principle, this reference architecture addresses the privacy controls mentioned above.

# 1 Introduction

The current state-of-the-art in federated identity management with respect to privacy protection comprises:

— Data minimization by limiting attribute release, i.e., identity authorities define policies that restrict which attributes may be released to a relying party, with or without explicit user consent [2];
— PII linkability prevention between relying parties, i.e., identifiers released to one relying party must not be linkable to any other relying party;
— General adherence to the common privacy and information security principles pertaining to identity management.

While these principles seem to be sufficient for users in a specific context, like an employee in her job duties, there are considerable privacy risks if federations work across contexts. Identity providers that have capabilities to track users may be targeted by over-curious employers, intelligence agencies exceeding their purpose or cyber criminals, leading to Orwellian surveillance. To mitigate these risks we propose a model that does not require users to trust that any single party is well behaved; rather at least two parties need to cooperate to disclose PII including usage data. The model shall be minimally invasive with respect to established protocols such as SAML WebSSO to minimize difficulties in adoption.

## 1.1 General Use Case

**Problem.** Distributed systems like networks and applications do not scale well if combined with local user management and authentication when access is restricted to trusted users.

**Forces**

— Complexity and effort for the enrolment for large user communities or across different organizations
— Scalability across domains, e.g., contractual relationships do not scale with isolated IdM
— Support of GRC using policy-controlled IAM
— Support of multiple and/or complex authentication protocols is difficult to manage for many relying parties

**Solution.** Delegate the management of identities, credentials and authentication methods to specialized entities (identity and attribute providers) in a brokered trust model.

**Related Pattern.** SSO Delegator [3]

## 1.2 Requirements

R1. Mitigate the **observability** threat. No federation entity shall be able to aggregate data about the usage of multiple services by users (principals), therefore being unable to deduct personal interests or behavior. (Compare to limited observability in the glossary.)

R2. Mitigate the **linkability** threat. Aggregation of PII used for different purposes or in different contexts is undesirable. That being so, two service providers processing data of a principal shall not be able to link that data without explicit user consent or the help of a designated opener. If complete pseudonymous authentication is not achievable, then at least those attributes that identify a user uniquely shall be pseudonymized. This applies, e.g., to the ubiquitous e-mail address. (Compare to limited linkability in the glossary.)

R3. Prevent the unauthorized **aggregation of attributes**. No federation entity shall be able to collect attributes beyond the specified purpose of a service and deduct personal information and behavior.

R4. Enable authorized **aggregation of attributes**. A service provider shall be able to request attributes from different attribute providers within the limits of the purpose of the service.

R5. Prevention of unauthorized **attribute polling**. A mechanism to prevent unauthorized discovery of attributes shall be provided.

R6. Prevention of **replay and reuse attacks**. An identity provider must limit each assertion to used only once at a specific service provider.

R7. **Consent handling.** The flow of releasing attributes should regard the processing of user consent, where explicit consent is appropriate.

R8. **No supreme deity.** Actors managing trust roots must not have access to either attributes or transaction data.

R9. Maximize **compatibility** with existing single-sign-on profiles to the extent that other requirements are not compromised:

- a. Feasible implementation effort. The model shall make use of existing profiles and implementations as far as reasonable.
- b. Feasible deployment effort. It shall be possible to use existing SAML service provider implementations within current configuration limits. For identity providers, that is desirable as well.

R10. Minimize the **release of attributes**. The identity authority, in its role as PII controller of a principal's identity information, must be assured that only those attributes deemed necessary for the purpose of the service are released to the relying party.

## 1.3 Privacy Implications of Current Federation Styles

Current WebSSO federations have two basic architectural styles: Peer-to-Peer (P2P or mesh) and Hub-and-Spoke[1]:

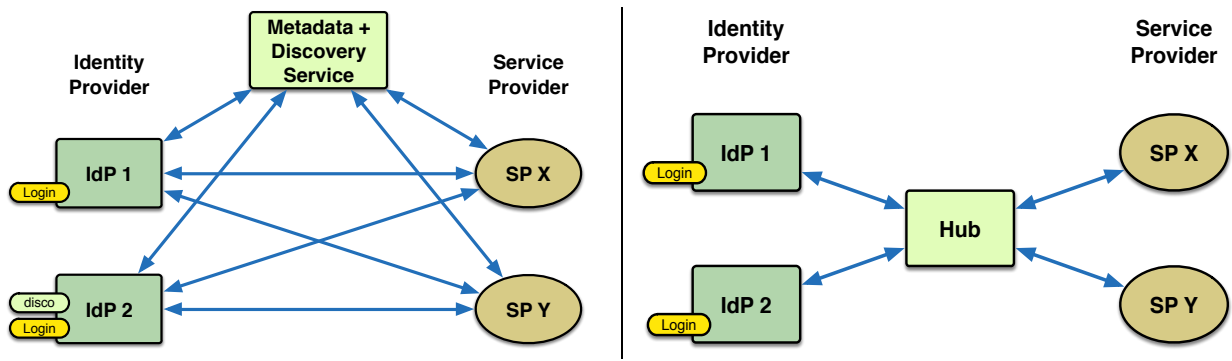

Fig. 1 Federation styles: left: P2P/mesh; right: Hub-and-Spoke

The privacy implication of both models is that the identity provider (IdP) can collect service usage profiles per user, thus violating R1. This is obvious in the P2P style because each IdP issues assertions for a specific service provider (SP). It is also implicit in the Hub-and-Spoke model, because the hub has a static 1:1 mapping for each service that might provide an attacker inside the IdP with enough information about the nature of services if analyzing data from many users, or impersonating a single user. In addition, the hub itself is violating R1 as well, as it is tracing user behavior. If the attributes were not encrypted with the SP's keys, then the hub would violate R3 as well.

---

[1] This model was outlined at the ACAMP 2012 session "Developing Federation Model Options". The transcript was available on 2013-Sep-15 at https://spaces.internet2.edu/display/ACAMPScribe2012/Thurs+3.30pm+ Salon4+-+Developing+Federation+Model+Options.

## 1.4 Privacy Implications of Attribute Release

Current best practice in large-scale federations, such as those from the higher education sector, is to deploy three main privacy controls: (1) a contractual agreement that binds the SP to limit the use of attributes[2], (2) the minimization of attributes by restricting the release of attributes per SP or SP affiliation, and (3) replacing the principal's unique identifier with a temporary or targeted ID.

Besides the lack of limited observability, the remaining problem in this practice is the sharing of attributes that are identifying, like e-mail addresses. Exchanging these attributes for pseudonymized ones would raise the bar for data protection. An implication is that this demands the implementation of additional processes in a federation, like the forwarding of messages from the SP to the user, and the opening of pseudonymous identities in case of a breach of contract or law.

## 1.5 Architectural Options for Identity Information Privacy

An overview on different approaches toward improving privacy in federations is given in the Kantara eGovernment WG [4]. The following list provides the pros and cons:

**(a) Identity Escrow**. The IdP/AP is taken out of the interaction with the RP, using cryptographic technologies based on group signatures like in IBM's Idemix and Microsoft's uProve. This provides an assertion to the RP without the IdP knowing to which RP they are asserting an identity.

> Pro: Technical control that satisfies the limited observability requirement.
> Con: (1) No implementation in mainstream products. Therefore, no capability of large-scale deployment in the near future; (2) Issues with other requirements (revocation, federation, IdP/AP business model); (3) Increased user dependency to trust technology provider due to increased complexity in crypto-technology.

---

[2] Terena: The Data protection Code of Conduct describes an approach to meet the requirements of the EU Data Protection Directive in federated identity management. Available on 2013-Sep-16 on
https://refeds.terena.org/index.php/Data_protection_coc

**(b) Late Binding**. Credential providers provide pseudonymous credentials to users, and RPs will bind attributes to those credentials.

This architecture was proposed by the Canadian Treasury Board for the CATS architecture. Their separation between credential service assurance and identity assurance implies that attributes are not stored with the IdP, but obtained by the RP [5] (p. 5).

>   Pro: Straightforward architecture that goes well with existing technology.
>
>   Con: (1) While mitigating R2, it does not fulfill R1. (2) The burden of binding attributes per RP diminishes the benefit of identity federation to some extent. (3) The collection of identifying attributes like name, residential and e-mail addresses is still likely; thus, R2 might not be fulfilled very well.

**(c) Proxy Pool.** Proxies that play RP to IdP/AP and IdP/AP to RP can significantly reduce the amount of data collection, if there are many of them.

>   Pro: (1) This is an evolutionary extension of existing technology, because proxies and gateways for IdM are a well-established technology and also part of the SAML specification.
>
>   Con: (1) Proxies provide no arbitrary choice, but rather one more actor in the equation, and add vulnerability. (2) Proxies would yield only a very limited constraint until the number of proxies is large; thus, it would be difficult to overcome the hen-and-egg problem.

**(d) User-based IdPs**. As proposed by IMI [6], the client would be the identity selector and could also hold the credentials locally. A similar concept has been proposed with personal authentication devices in [7].

>   Pro: This architecture has strong security and privacy controls.
>
>   Con: (1) Deployment problem because it is difficult to enhance web browsers and (2) with PKI-based credentials there is still the tracking issue with OCSP responders. (3) CardSpace as the only implementation from a large vendor was declared "feature complete" by Microsoft in 2011, which was a big setback for the IMI community. Tying identities to desktops was a failure. (4) Experience with the "Neuer Personalausweis," the German national identity card, showed that complex deployment leads to the growth of cloud services that offload some deployment issues, but violate R1 in turn.

**(e) Organizational Controls.** The FI Workgroup of Kantara proposed two controls: (1) Regulation will mitigate the risk with preventive organizational safeguards, liability and legal enforcement. (2) Providing choice with IdPs and enforcing transparent and efficient markets will drive providers to comply with privacy requirements.

Pro: As an analogy this works well in the financial services industry, where banks have a panoptical view on financial transactions of their clients, but are still trusted due to the regulation.

Con: (1) Abuse would be possible without collusion with another actor, because this is not a technical control. (2) Consumers, if privacy-aware at all, typically do not act to improve better privacy [8].

**(f) The Privacy-enhanced FIM Architecture.** The model introduced below enhances the hub-and-spoke model by offering technical controls that enforce limited observability and enable pseudonymous authentication.

Pro: It proposes reasonably strong technical controls.

Con: Despite not requiring a lot of new technology, as option (a) does, it is still not fully compatible with existing implementations; therefore, it will not run out of the box with existing products.

**Design Rationale:**

The architectural options (a), (c) and (d) will not be feasible in the next couple of years, if ever. Option (b) has a weak value proposition because attributes cannot be reused, and does not prevent at all that identifying attributes will be collected by different relying parties and later on help to link up PII between different services. Option (e) reflects current practice, but does not fulfill higher privacy requirements, and is particularly vulnerable to state-sponsored surveillance and commercial espionage because the extraction of traffic data is the first step in analyzing potential targets.

Decision: (f) is selected as the most feasible option to provide enhanced technical privacy controls.

## 1.6 The Privacy-enhanced FIM Architecture (PE-FIM)

This model proposes an approach to federated identity management (FIM) that is privacy-friendly with respect to the requirements defined above. It is based on a 3-tier architecture that is an extended hub-and-spoke model with privacy by design principles applied to it. The hub is called the service broker (SB) in this model.

### *1.6.1 Privacy by Design Approach*

Privacy by design (PbD) is a high-level construct devised by privacy commissioners to raise the bar for data protection in comparison to Fair Information Practice Principles (see [9] p. 177). While the abstraction levels of PbD and this FIM architecture prevent the direct inherence of specific technical controls, this architecture is in line with PbD principles because it:

   a) Is going beyond the common legal-oriented approach to privacy as this design emphasizes embedded technical controls (see principle 3 in [10]);
   b) Separates the processing of the data subject's attributes across different actors;
   c) Is preventive, because data that is not available to an actor does not need protection.

### *1.6.2 PE-FIM High-level Architecture*

The very outset of the PE-FIM model is the introduction of a secure pseudonymous channel to support requirements R1, R2 and R3. The desired property of this bidirectional channel is that an IdP and an SP, or two SPs, can communicate about a principal, where (a) the SPs are pseudonymous to the IdPs, (b) the principal is pseudonymous to the SPs and (c) the IdP's and SP's identities are vouched for by the certificate authority.

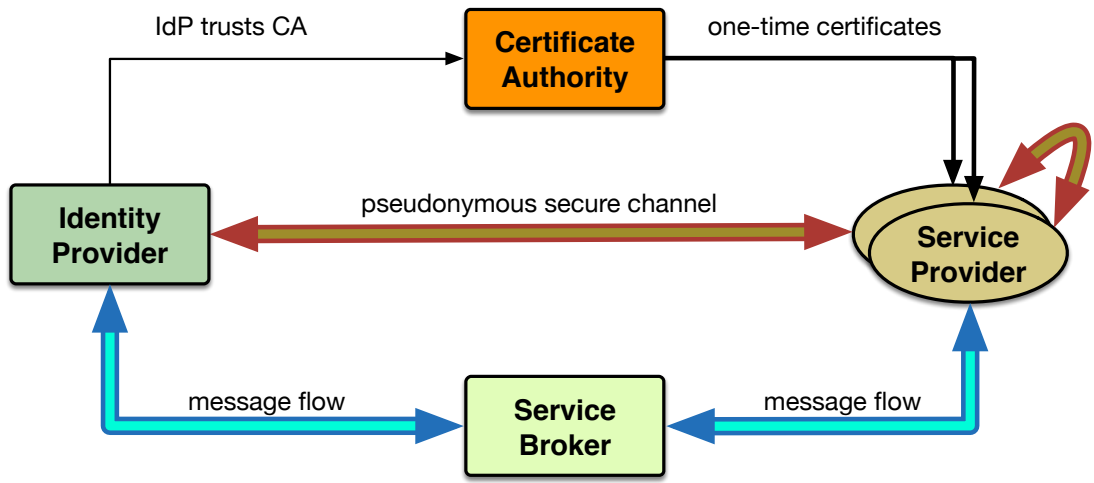

Fig. 2 High-level Architecture

It is assumed, but not show in the picture above, that trust has been established between SP-SB and SB-IdP, using certificates or other means.

The core constructs of the proposed model are:

a) A pseudonymous secure channel, which can be used for several purposes:
    i. Transmit assertions from an IdP to an SP;
    ii. Transmit pseudonymous data about a principal between SPs;
    iii. Transmit security alerts or operations-related messages from the SP to the IdP;
    iv. Transmit application-level messages from the SP to the principal, relayed by the IdP using SMTP or another messaging protocol.
b) The secure pseudonymous channel is implemented using a mixture of brokered trust and end-to-end encryption.
    i. All messages are relayed via the service broker (SB);
    ii. The "payload part" of messages is encrypted for the final recipient; the rest is unencrypted to be processed by the SB. Payloads would be attributes for the assertions, subject and body for other messages;
    iii. Targeted IDs are never encrypted;
    iv. Users are pseudonymized for SPs by sending targeted identifiers;
    v. Users are anonymized for SBs by sending encrypted payloads.
d) Each message for an SP is encrypted using a one-time key for pseudonymization of the SP;
e) Reference IDs are transformed into targeted IDs in a 2-step process: (i) from the IdP to the SB, and (ii) from the SB to the SP.
f) Targeted IDs can be used for messaging, payment and delivery and shall make the release of identifying attributes like email addresses and credit card numbers superfluous.
g) The model needs to have several similar SPs with respect to transaction volume, requested attributes and bindings. They have to be grouped together by an SB to appear as a single entity toward an IdP. Proxying an SP in a 1:1 fashion would be too transparent to offer privacy protection.
h)

### 1.6.3 Trust Model

This trust model excludes:
— Availability and business continuity aspects;
— Application-level confidentiality and integrity aspects, e.g., for content that is owned by the relying party or third parties.
— Protection against attacks on components that are not related to identity management.

|   | Asserting Actor | Relying Actor | Trust |
|---|---|---|---|
| 1 | IdP | SP | Entity identification (including identity proofing and verification) and entity authentication |
| 2 | User | IdP | Secure the authentication process (e.g., protect authentication token and end-user device) |
| 3 | IdP | User | IdP will not falsify assertions to impersonate the user; Proper handling of PII given away in the registration process |
| 4 | RP | User | Proper handling of PII that was released with the identity assertion (might be indirectly via the IdP) |
| 5 | Certificate Authority | RP, IdP | Provide a trust anchor |
| 6 | IdP | SB | Relay the attribute release policy on SPs |
| 7 | IdP | SB | Obtain proper consent from user for an attribute release (optional) |

### *1.6.4 Use cases*

The general model is applied to two common use cases:

a) The WebSSO use case. Principals accessing an SP with a web browser authenticate with an IdP, which may also provide attributes. The SAML WebSSO profile [11] is a well-known instance of that model. It provides several protocol bindings, but for simplicity this model builds upon the widely deployed front-channel protocol to convey attributes to the SP.

b) The Web Service use case requires the principal to authenticate directly with the SP using a pseudonymous X.509 certificate trusted in the federation. The SP will then query attributes from an attribute authority.

## 1.7 Applying the PE-FIM Architecture to the WebSSO Use Case

This section explains the model by starting with the basic 2-tier model, adding the third tier and then the privacy-enhancing service broker. WebSSO is widely deployed in both mesh-type and hub-and-spoke federations using the SAML WebSSO profile that is again profiled by SAML2Int[3]. Terminology and use case details are aligned with SAML WebSSO for this use case, because SAML seems to be the preferred candidate to implement the model, given its broad adoption and extensibility.

### *1.7.1 Basic Authentication Delegation*

Fig. 3 shows the interactions for authentication in a multi-IdP federation using a front-channel authentication request protocol.

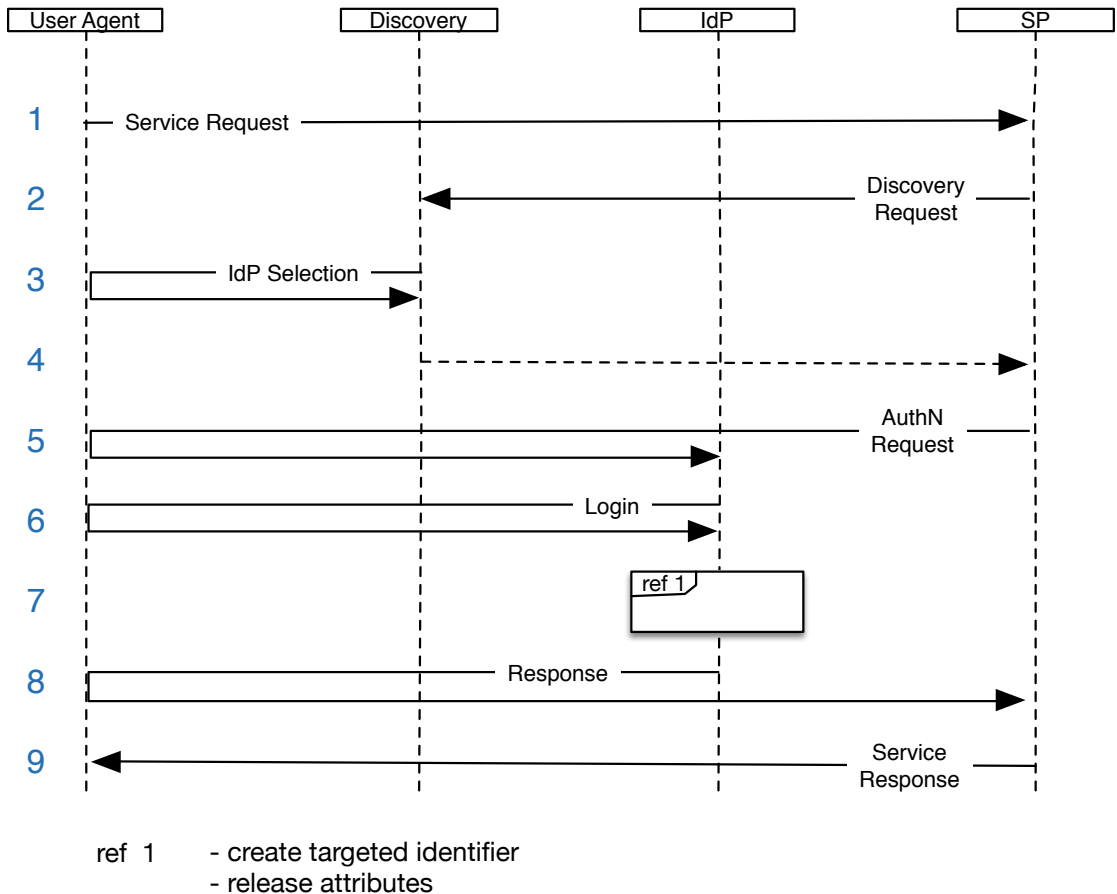

ref 1    - create targeted identifier
              - release attributes

**Fig. 3 Basic WebSSO: authentication delegation in a 2-tier (mesh) model**

---

[3] SAML2int (http://saml2int.org) is an SAML profile deployed in many research and higher education federations, such as InCommon or SWITCH.

Hub-and-Spoke federations extend this model by splitting the federation on a protocol level into two federations, with all IdPs on one and all SPs on the other side of the hub. The interaction pattern can be derived from the basic WebSSO pattern described above by applying it to either side of the hub and chaining them together. This process is specified in [11] section 3.4.1.5.

In step 7 the IdP creates a response including an identifier and possibly other attributes of the principal. To limit linkability the identifier is either a temporary ID or a targeted ID specific to the IdP-SP pair. In either case the identifier is based on a pseudo-random value that has no discernible correspondence with the subject's actual identifier. (Compare to [11] section 8.3.7.)

The WebSSO use case is implemented in various protocols such as SAML, Open ID Connect, OpenID 2.0, and WS-Federation Passive Requestor Profile. Its different protocol options, such as pushing assertions via the front channel, or querying the assertion via a back channel, can be ignored for the purpose of this model.

### *1.7.2 3-tier Model with a Service Broker*

The key responsibilities of the SB introduced here are the relaying of messages and mapping of targeted IDs to fulfill R1 (limited linkability).

Fig. 4 provides an informal exposition of the proposed model's key components and their relationships.

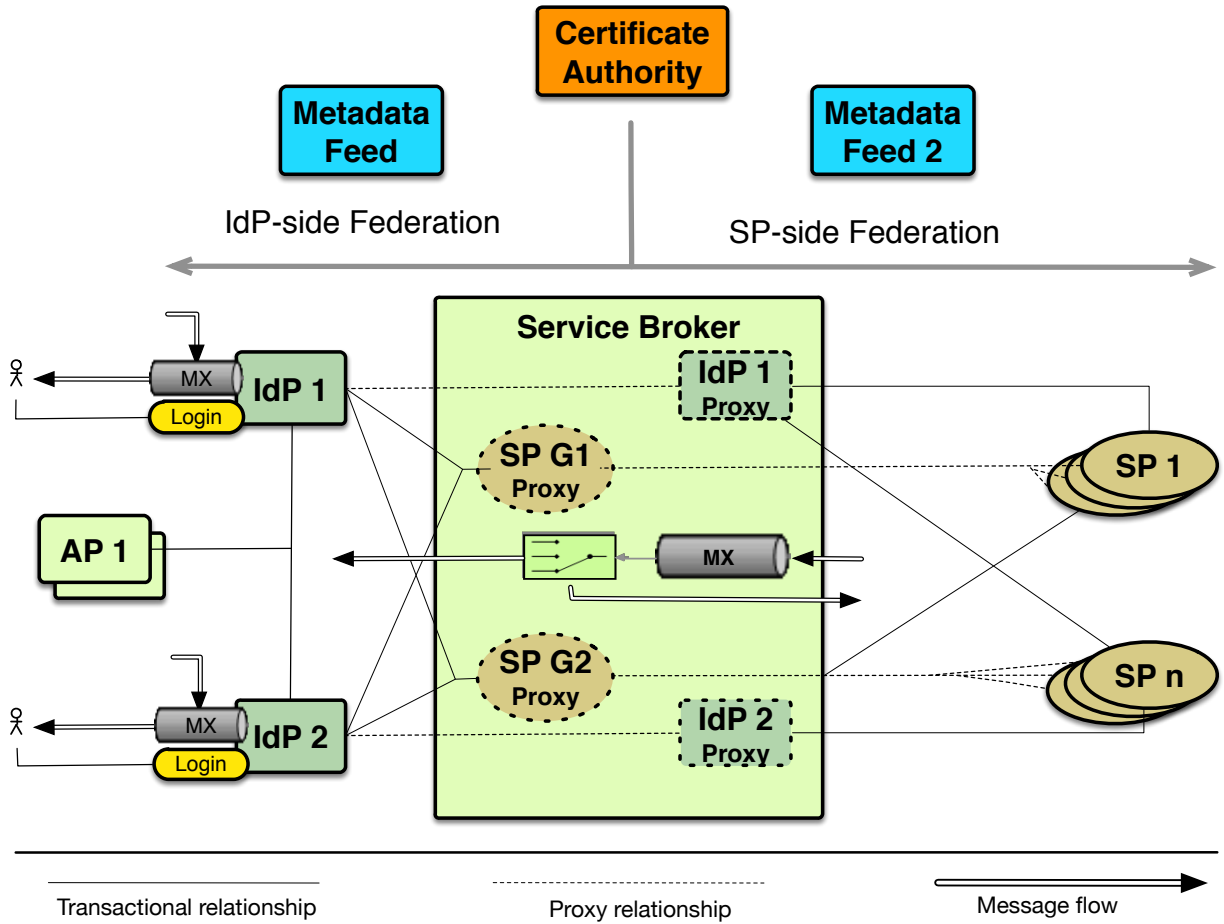

Fig. 4 High-level layout of the PEFIM model (WebSSO use case)

The structural features of this model are:
— Each IdP has a 1:1 representation as an IdP proxy in the SB
— All SPs with the same characteristics (required attributes, protocols and bindings) share an SP proxy in the SB.
— The SB provides a messaging service (SMTP, XMPP or others) including the mapping of targeted addresses.
— Attributes are encrypted end-to-end between the AP and SP (or IdP and SP). For this purpose an SP uses dynamic encryption keys that are provided by the certificate authority.
— Metadata feeds provide static configuration and trust information for each sub-federation.

Identifying information is made available to SPs to a minimal extent, by using targeted identifiers and providing services to replace identifying attributes such as email addresses and credit card numbers. Tracing information is hidden from IdPs and APs because no SP is identified to them.

Attributes (except identifiers) are unavailable to service brokers because they are encrypted for SPs.

A prerequisite of this model is an organizational separation of the certificate authority and service broker.

### 1.7.3 PE Model Flows

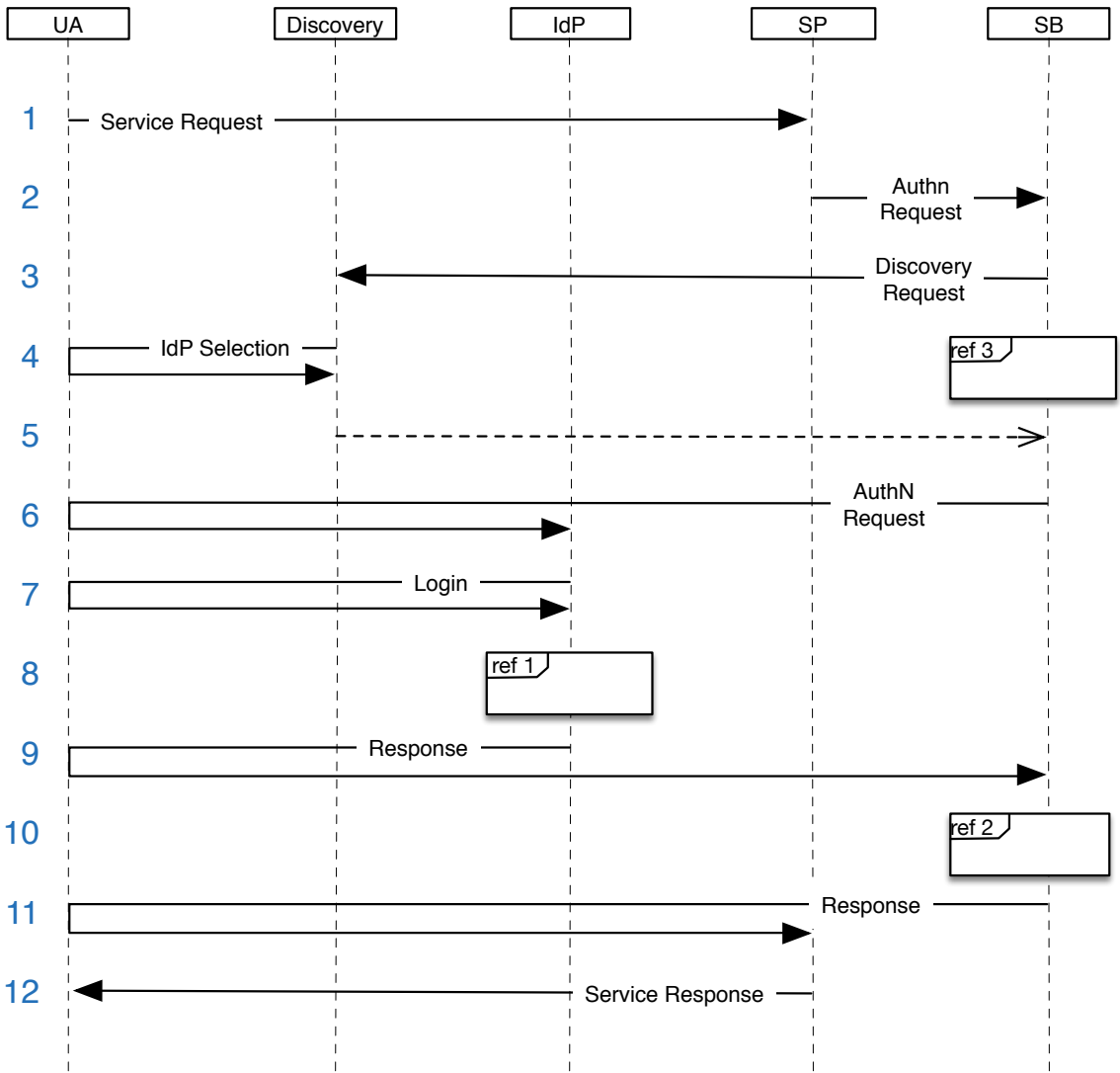

| ref 1 | - create targeted identifier for SB<br>- encrypt attributes for SP |
| --- | --- |
| ref 2 | - create targeted identifier for SP |
| ref 3 | - optional: obtain user consent |

Fig. 5 3-tier model with a service broker

### *1.7.4 Attribute Encryption with One-time Keys [R3]*

Attribute encryption in typical SAML federations uses static encryption keys provided with metadata, or configuration data in small-scale deployments. Such a static encryption key would, however, identify the SP to the IdP and, thus, violate R1.

To encrypt attributes without violating R1 a one-time encryption key must be used that changes per transaction. The key is generated by the SP and signed by the certificate authority.

Thus, any instance forwarding the attributes from the AP to the SP (i.e., IdP and SB) cannot read the contents. The encryption certificate is included in the authentication request and needs to be validated by the IdP or AP.

NOTE: Keys for signing authentication requests and encrypting messages for IdPs should remain static. They should be provided by metadata.

### *1.7.5 Targeted IDs*

In this 3-tier model, targeted IDs are used between the IdP and the SP proxy that is part of the SB, and between the IdP proxy (again part of the SB) and the SP. It is the SB's responsibility to map the targeted IDs from the IdP to those at the SP, and vice versa.

| ID Type | Generated/stored at | Function |
|---|---|---|
| Reference ID | IdP | |
| Targeted $ID_1$ | IdP | One-way function (Reference ID, SB-ID) |
| Targeted $ID_2$ | SB | One-way function (Targeted $ID_1$, SP-ID) |

### *1.7.6 Message Delivery by IdPs*

Targeted IDs can be used as the address in messages, but need to be converted to a service-specific format such as an RFC822 e-mail address. The assumption here is that the IdP is capable of providing a message forwarding service such as SMTP and XMPP (Jabber) instant messaging. In the case of an e-mail address the targeted ID can be used as the user part, with the SMTP relay of the SB as the domain. The address rewriting would look like this for e-mail/SMTP:

|  | IdP | SB | SP |
|---|---|---|---|
| E-mail address | user@homeorg.tld | $TID_1$@mail.idp.tld | $TID_2$@mail.sb.tld |
| MTA | smpt.idp.tld | smtp.sb.tld | |
| Mapping | $TID_1$ -> user@homeorg.tld | $TID_2$ -> $TID_1$ | |

Table 1: Messaging functions at SB and IdP

The e-mail addresses used by SBs and SPs are derived using a fixed schema. That is the Targeted ID 1 ($TID_1$) issued by the IdP side to forward messages between the SB and the IdP, and the Targeted ID 2 ($TID_2$) issued by the SB to forward messages between the SP and the SB. The same pattern can be applied to other addresses such as XMPP or SMS.

This pattern can be extended to deliver messages from the SP to the SB and IdP by augmenting the addressing schema. The use cases for this are consent management, reporting of security alerts (like the abuse of an account at the SP), aggregation of usage statistics, and updates of attributes at the SP.

### *1.7.7 Targeted IDs to Link User Data Between SPs*

There are use cases where a user's data sets with two SPs need to be linked up. A typical case is a communication or delivery service having the role of an SP. Examples might be yellow mail, paid electronic delivery services such as De-Mail, or a payment service. On that condition an explicit link between the two SPs needs to be established. The privacy-preserving technique is to use the SB again as a de-identifying broker by creating a pair of unique identifiers between SP1 and SP2 (or a single ID for a one-way link):

| ID Type | Identifier | Generated/ stored at | Function |
|---|---|---|---|
| Reference ID | | IdP | |
| Targeted $ID_1$ | $TID_1$ | IdP | One-way function (Reference ID, SB-ID) |
| Targeted $ID_2$ | $TID_2$(SP1) | SB | One-way function (Targeted $ID_1$, SP1-ID) |
| Targeted $ID_2$ | $TID_2$(SP2) | SB | One-way function (Targeted $ID_1$, SP2-ID) |
| Targeted $ID_3$ | $TID_3$(SP1, SP2) | SB | One-way function (Targeted $ID_1$, SP1-ID, SP2-ID) |
| Targeted $ID_3$ | $TID_3$(SP2, SP1) | SB | One-way function (Targeted $ID_1$, SP2-ID, SP1- |

| | | | ID) |

Table 2: Pseudonymous delivery to another SP (exemplified with physical shipment)

One-way function in this context means that there is no way to compute or guess the value of the input IDs from the value of the targeted ID. The issuing entity has to keep a mapping table to make that relationship reversible.

The structure described in Table 2 supports the following flow:
a) SP1 needs to send a message to SP2 referencing a user;
b) SP1 obtains $TID_3$ (SP1, SP2) from the SB using an appropriate consent flow;
c) SP1 delivers the message to SP2 using $TID_3$ (SP1, SP2) as the identifier;
d) SP2 calls SB to convert $TID_3$ (SP1, SP2) to $TID_2$ (SP2);
e) SP2 can send a return message using the reverse process.

$TID_3$ is permanent, and a link once given cannot be revoked.

### 1.7.8 User Consent

User consent is a function to inform the user about the intended release of attributes to an SP and ask for his or her permission. Consent is posited among others by Laws of Identity [12] and the European Data Protection Directive [1].

#### 1.7.8.1 Consent Feasibility

Consent is a controversial topic, in particular because

(a) Study participants remember little about the contents of consent, even if a paper document on health-related matters was explained orally, and

(b) The literal requirement of specific and informed consent can lead to interactions with poor usability.

Solove [13] writes:

"The EU has a more paternalistic approach to data processing [..] EU privacy law has a self-management component, but it requires a much more stringent and explicit form of consent than U.S. privacy law. The difficulty with the EU approach is that data collection, use, and disclosure are rarely inherently good or bad."

"People want some privacy self-management, just not too much. Privacy law needs to find a way to deliver partial privacy self-management." [loco citato]

Specifically, as the UK Information Commissioner puts it, "The collection and use of personal information is often essential to provide the service or carry out the transaction that the user has requested. In such cases choice is not an issue, because the user cannot expect to receive what he or she has asked for unless any necessary processing of personal information takes place." ([14] p. 12)

However, even if consent is seen as the last resort to support a user's privacy requirements, there needs to be an option to collect consent without introducing a new data aggregation problem.

### 1.7.8.2 Consent Service

User consent may have several forms, as prescribed in [1] Art. (30):
1. It may be granted out-of-band (e.g., a customer signing Terms of Use on paper for a bank account that includes a consent to process identity attributes for a specific purpose; waived by legal requirement or inferred from an employment contract);
2. Consent might be acquired up front for any transaction, e.g., when an app is installed on a mobile device;
3. Consent may be asked in the context of a transaction.

From these descriptions and common administrative practice one can derive the following requirements for a consent service:
a) Consent must be stored and made available to IdPs for attribute release;
b) Consent must be captured using the different processes as described above;
c) Users must be provided with an interface to review and revoke consent;
d) Consent shall be managed and stored by service brokers, because this is the only location in a federation where R1 would not be jeopardized. The reason is that consent is given by a user targeting a specific SP. The IdP cannot maintain such a link, because the IdP must not identify individual SPs. Of course it cannot be stored with an SP, because consent shall protect the user from releasing attributes to the SP.

### 1.7.9 Check for Requirement Fulfillment

This section tries to prove that the design fulfills the stipulated requirements. To start with, the table below shows the exposure of privacy-related data items to different federation actors. Then each requirement and related control is discussed. This informal analysis is restricted to the most important aspects.

| Entity | Client IP Address | SP Identity | User Identity | Signing Key CA | Encryption Key SP | Required Attributes |
|---|---|---|---|---|---|---|
| CA | | | | X | certify | |
| IDP | X | only SB | X | | public | |
| SB/CS[4] | X | X | pseudonym | | | |
| IdP-DS | X | X | | | | grouped |
| SP | X | self | pseudonym | | private | |

Table 3 Exposure of data elements to federation actors

### 1.7.9.1 Requirement check – R1

R1 requires that the SP not be identified to the IdP. This implies that an authentication request does not contain any element identifying an SP. This is fulfilled, because authentication requests are being issued from SBs, with only this additional data from the SP:

a) Encryption certificate. It is a one-time certificate not containing the SP's name or identifier;
b) Proxy SP identification (SAML EntityID) and endpoint addresses. SPs are not mapped 1:1, but several SPs with the same set of requested attributes, NameID policy and bindings share one proxy SP.

As a consequence, larger numbers of SPs work better in diffusing links between principals and SPs.

The main unresolved or difficult problems are those of IP address tracking, physical delivery and payment, which are conventionally out of scope for federated identity management solutions.

**IP address tracking.** For some users the relationship between a fixed IPv4 address and the user identity can lead to tracking by IdPs. Pseudonymization solutions like dynamic addresses, VPN connections and overlay networks may help to mitigate this problem. A potential solution would require the SB to act as a forward proxy, but viable business and operation models would have to be found.

**Physical delivery.** Private address forwarding has been very recently proposed to the US Postal Service[5], but it needs to be seen whether a viable business model can be established for such a service.

**Payment.** Various pseudonymous payment protocols have been proposed and introduced, e.g., by [15], but no one seems to have gained broad acceptance. For example, PayPal has stopped offering virtual credit cards [16], which could have filled this niche[6]. This failure to establish this privacy-improving feature might be owed to a higher level of trust in banks, based on the hypothesis in [17] that institutional characteristics (highly regulated industry) influence trusting behavior and lessen the demand for privacy-preserving options.

---

[4] Consent Service, can be part of the SB.
[5] http://www.prc.gov/Docs/87/87900/Order%20No.%201838.pdf

### *1.7.9.2 Requirement check – R2*

Using SP-specific identifiers is a standard best practice and implemented in SAML. The additional challenge is to apply this to other attributes as well. The solution proposed here is two-fold:

a) For communication addresses the mapping scheme of targeted IDs is used. The prevalent use case is an e-mail address. This is provided by offering a targeted e-mail address for each principal to each SP. The same principle could be used for XMPP, SMS, telephone and physical mail, but the infrastructure cost is substantial for the latter options.
b) For other attributes, late binding or self-provisioned attributes are used. For example, a user could choose a display name that might or might not relate to other SPs.

### *1.7.9.3 Requirement check – R3*

The aggregation of attributes by the SB (or any other party) is inhibited by the end-to-end encryption of attributes.

NOTE: The unlawful proliferation of attributes by SPs can only be mitigated by organizational controls.

### *1.7.9.4 Requirement check – R4*

There is no need to establish a completely new technology stack for identity federations to achieve WebSSO functionality. The most widely deployed protocols can be enhanced to fit to the model. The considerations in the case of SAML are:

i) The public encryption key is passed as an extension element in the <samlp:authnRequest> in a <KeyDescriptor> element analog to the specification in [18] section 2.4.1.1. IdPs and APs must consume the encryption key from this extension.
j) Validation of dynamic encryption certificates: Large-scale SAML federations typically use static encryption keys validated against metadata [19], instead of X.509 path validation.

---

[6] PayPal removed the web page offering a free virtual debit card:
https://www.paypal.com/kh/webapps/mpp/security/general-freetools.

The PE-FIM model requires the encryption certificate to be consumed from the request and path-validated using X.509 processing rules.

k) Assertions are only encrypted partially. NameID elements need to stay unencrypted for processing by the SB. All other attributes must be encrypted. SAML provides encryption by attribute, but not all products do support this[7] and there is some overhead due to the use of XML Encryption per element.

l) As there is no option in SAML to encrypt only the attribute statement, it would be possible to split the authentication statement and the attribute statement into two assertions: the first one unencrypted and the second one encrypted. SAML allows a response to contain multiple assertions, but it is uncommon and might not be supported by all implementations. However, there is a fallback path. If a federation entity does not produce or consume multiple assertions per response, a different protocol pattern can be configured to deliver the authentication statement in the first response and perform an attribute query in a second process step.

### 1.7.9.5 Requirement check – R5

Authentication requests and attribute queries need to be authenticated and require a targeted ID. Therefore, attribute polling is prevented.

### 1.7.9.6 Requirement check – R6

Replay and reuse attack mitigation is based on standard SAML controls, such as transport security, one-time use of assertions and artifacts, pairing of requests and responses, and use of audience restrictions.

### 1.7.9.7 Requirement check – R7

Consent handling is provided by the SB with preserving R1. A UI implication is that the SB cannot display clear display names of the user, but that could be improved by rendering the consent service in an iFrame from the IdP.

---

[7] e.g. ADFS does not support attribute encryption: http://social.msdn.microsoft.com/Forums/vstudio/en-US/a63e31cb-a109-4e99-8538-2eb084ed3827/support-for-encrypted-attributes-in-saml-20?forum=Geneva

### 1.7.9.8 Requirement check – R8

A prerequisite of this model is the organizational control, wherein the FO who controls the certificate authority must be organizationally separated from all other federation entities, in particular from the SB.

### 1.7.9.9 Requirement check – R9

The most widely adopted SAML WebSSO interoperability profile for federated identity management is the Kantara eGov profile and its derivate deployment profiles such as the Saml2Int[8].

Extending from this profile, the technical impact is located in two areas:
a) The generation and validation of encryption certificates, and
b) The requirement of a service broker, which extends the concept of a SAML IdP proxy with identifier mapping and consent handling.

The critical point is to be able to integrate service providers into a federation with as little pain as possible. It needs to be investigated if a) could be resolved for standard SAML-enabled service providers using a proxy component.

### 1.7.9.10     Requirement check – R10

The use case above does not cover the onboarding process for service providers. However, the concept of entity categories, which describe the purpose of a class of services, allows abstracting the attribute release policy from a specific SP to a group of services represented by a service broker. This would imply that the identity authority delegates the responsibility for the attribute release to the service broker.

## 1.7.10     IdP Discovery

Both SP-embedded and centralized discovery services may be used to furnish the selection of the user's IdP using home organization, location and other properties.

---

[8] http://kantarainitiative.org/confluence/display/fiwg/SAML+Interoperability+and+Deployment+Profiles

## 1.8 Web Service Use Case (SOAP)

### *1.8.1 Introduction*

Web browsers are passive clients in the WebSSO case with respect to authentication. They do not store credentials and have no initial knowledge about the IdP. Opposed to this, WS clients are active, because they do store credentials and know the IdP's address. This results in a much larger variety of authentication scenarios.

The following specifications and protocols are available for SOAP:

1. WS-Security is the security building block within the WS-* protocol stack for the basic bilateral use case. It standardizes how security information is added to SOAP messages. WS-Security headers contain security tokens such as Kerberos tickets, SAML assertions or userids/password pairs to convey the client's identity. Messages may be protected by signatures or TransportBinding. It is state-less and can be quite secure if implemented properly[9][20-23]. However, interoperability and delegation can be problematic because of the many options available in the specification and it requires parties to exchange keys and agree on the security token format out-of-band. WS-Interoperability specifies the Basic Security Profile[24] to resolve some interoperability issues at the protocol level.
2. WS-SecureConversation establishes a secure transport between client and server, providing primarily better performance at the price of added complexity.
3. WS-Trust introduces more generic scenarios by brokering trust and converting security token formats. The Security Token Service (STS) can have a role comparable to an IdP in the generic model, issuing security tokens. Or it can assume the role of a validator, attesting a claim made by the requesting WS client.
4. Subject confirmation methods (bearer, holder of key, sender vouches) attest the binding between the security token and the SOAP message, which may be combined with different options from above.
5. The various options may be configured locally into a product, or use WS-SecurityPolicy as metadata.

---

[9] XML-Signature and XML-Encryption need protection against known vulnerabilities like XML signature wrapping and other attack. And, of course, it is crucial avoiding the rich set of possibilities to do XML security in the wrong way and doing right decisions with respect to the flexibility, complexity and incomplete documentation of WS-*.

Not all of these combinations are feasible in a federated environment. Assuming that attributes need to be assured beyond the X.509 subject name, SAML security tokens containing attribute assertions are required.

As an alternative just an identifier could be included in the message, and Backend Authentication Exchange (BAE) [25] could fetch the attributes. While this is not specified within the WS-* architecture, it would reduce the message size if attribute assertions are cached.

SAML tokens have the additional advantage that subject confirmations and audience restrictions provide increased security.

The model in the following section describes a SOAP messaging using WS-Trust with SAML security.

### 1.8.2 WS-Trust with SAML Tokens (Draft Model)

Assumption: SAML assertions are being used as security token. The audience restriction is enabled to assure that SAML assertions cannot be reused for unintended web services.

Requirement: The audience restriction for the WS server (SP) needs to be hidden from the STS-IdP.

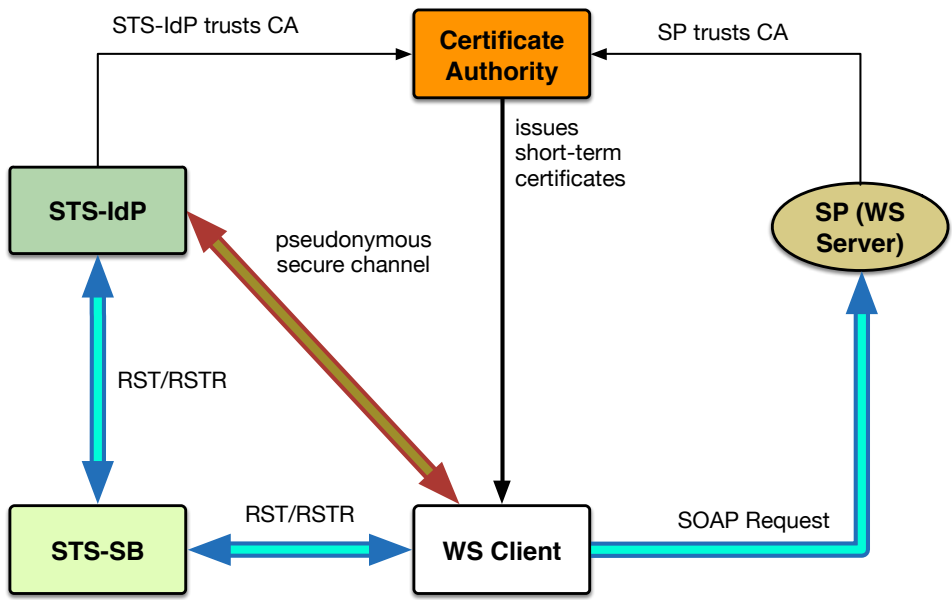

Fig. 6 STS issuing security token

The privacy-enhanced model described in Fig. 6 can be used with bearer and holder-of-key subject confirmations. It is applicable to both WS-Security with TransportBinding and message signatures.

The WS client obtains a short-term certificate to be used for authenticating the RST message, decrypting the SAML attribute assertion, and authenticate to the SP. The flow has these main steps:

1. The WS client requests a security token from the STS-SB (RST). The STS-SB verifies the authentication in order to conform to R5.
2. The STS-SB proxies the request to the STS-IdP. The RST does not contain information about the SP, supporting requirement R1.
3. SAML assertions issued by the STS-IdP contain a targeted ID for the STS-SB, and attributes encrypted with the WS client's key, supporting requirement R3.
4. The STS-SB proxies the request to the WS client, replacing the targeted ID with the targeted ID for the SP (fulfilling R2), and creating an audience restriction for the SP (-> R6)
5. The WS client unencrypts the attributes, and then sends the SOAP message with the SAML security token that contains a targeted identifier and the requested attributes. Confidentiality and integrity of the SOAP message can be protected by standard WS-Security methods.

Impact on WS-* standards. There is some new functionality required by the WS client to work with short-term certificates and decrypt the SAML attribute assertion. The STP-IdP needs to be adapted for encrypting the SAML assertion with the short-term certificate. The communication between WS client and WS server is not impacted by the enhancement. The additions to the client could be encapsulated into an adaptor for the RTS/RSTR message exchange. There is a moderate impact on the STS and WS client, providing a compromise with regard to requirement R9.

# 2  Conclusion

The reference architecture for privacy-enhanced FIM addresses requirements that prevent surveillance and data mining across identity and service providers, most significantly due to the technical inhibition of collecting user behavior data. It claims to deal with a problem that has not been solved in commercial projects so far. The model features a design that allows an evolutionary migration from existing technology stacks, thus increasing the probability for market adoption. A detailed model has been drafted for the SAML WebSSO use case, and a draw-up shown for SOAP web services.

It is thus complementing approaches that employ PET based on advanced cryptographic technologies, such as blind signatures and zero-knowledge proofs.

# 3 Glossary

## 3.1 Abbreviations

| | |
|---|---|
| *AP* | Attribute Provider |
| *FIM* | Federated Identity Management |
| *IA* | Identity Authority |
| *IdP* | Identity Provider |
| *PET* | Privacy Enhanced Technology |
| *PII* | Personally Identifiable Information |
| *PKI* | Public Key Infrastructure |
| *RP* | Relying Party |
| *SAML* | XML-based framework standardized by the OASIS Security Service Committee to communicate the authentication, attributes and privileges of users |
| *SB* | Service Broker |
| *SP* | Service Provider |
| *UA* | User Agent (a web browser in the WebSSO use case) |
| *WebSSO* | Single Sign On (SSO) for clients using a web browser |

## 3.2 Definitions

Definitions formatted in italics are taken literally from the cited standards.

### Assertion

Trustworthy identity information produced by an identity provider to be consumed by a service provider.
Examples: SAML authentication assertion, SAML attribute assertion, WS-Trust security token, OAUTH token, Kerberos ticket.

### Attribute

*Characteristic or property of an entity that can be used to describe its state, appearance, or other aspects.*

*NOTE: The primary function of the concept of an attribute is to be a particular, well-defined aspect of the description of an entity in an identity management system. The values of attributes in an identity together describe the entity in a domain.*

*EXAMPLE: An entity type, address information, telephone number, a privilege, a MAC address, a domain name are possible attributes. [ISO/IEC 24760-1]*

*Attributes may consist of something*
- *the entity is or has,*
- *the entity has chosen,*
- *the entity has been assigned. [ITU-T X.1252]*

### Attribute Provider

Technical component that maintains attributes about a principal and makes them available in the context of an authentication or service provisioning to a service provider.

NOTE: The term is a synonym to attribute authority as defined in [SAML Glossary].

### Attribute Release Policy

Defines the set of attributes that may be disclosed to a relying party. It can be defined per relying party, based on the data minimization principle, or per user, where additional restrictions may be defined by consent.

### Authentication

Corroboration of claimed information (e.g., a set of attributes) with a specified, or understood, level of confidence in the context of an electronic communication.

NOTE: In the context of this architecture, authentication stands for both unilateral entity authentication with the goal to corroborate the claimed identity of a subject to establish a session or security context, and for message authentication. While both forms of authentication differ in protocol and timing behavior, most aspects of identity management apply to both forms.

### Connection Data

Data that reveals where, when and for how long a user accessed a service, without knowing about the contents of the connection.

### Control

*Means of managing risk, including policies, procedures, guidelines, practices or organizational structures, which can be administrative, technical, management, or legal in nature.*

*NOTE: Control is also used as a synonym for safeguard or countermeasure. [ISO/IEC 27000]*

### Credential

*Set of data presented as evidence of a claimed identity and/or entitlements. [ITU-T X.1252], [ISO/IEC 29115]*

NOTE 1: ISO/IEC 29115 Annex B describes properties of a credential in detail.

NOTE 2: This definition implies that the physical token containing a credential is not the credential itself.

### Designated Opener

Group of entities that can re-identify the principal by linking attributes and identifiers.

NOTE 1: This definition corresponds to that from ISO/IEC 29191 into a similar use case.

NOTE 2: The designated opener is a role that needs to be assumed by different actors depending on which actor needs to re-identify a principal.

### Entity

*Item inside or outside an information and communication technology system, such as a person, an organization, a device, a subsystem, or a group of such items that has recognizably distinct existence. [ISO/IEC 24760-1]*

NOTE: It can be a dog as well, of course.

### Federated Identity Management

The concept of cross-organizational cooperation to manage and use electronic identities.

### Federation

Association of actors with a common interest in organizing trustworthy electronic communication that joined for this purpose.

NOTE: The federation can be restricted to a certain market, common interest or geographic region. See also 'Identity Federation.'

### Federation Entity

One of identity provider, service broker or service provider.

### Federation Operator

Entity providing day-to-day operational management and support for a federation.

NOTE: The federation operator is typically authorized to enter into binding contracts and to provide support for federation services. The federation operator manages federation members and controls a technical trust anchor that is the base for other cryptographic trust relationships. That data is usually published as a metadata service.

### Front Channel

*Front channel refers to the "communications channel" that can be effected between two HTTP-speaking servers by employing "HTTP redirect" messages and thus passing messages to each other via a user agent, e.g. a web browser, or any other HTTP client [RFC2616]. See also back channel. [SAML2Core]*

### Identifier

*Identity information that unambiguously distinguishes one entity from another one in a given domain.*

*NOTE 1: An identifier may be suitable for use outside the domain.*

*NOTE 2: An identifier may be an attribute with an assigned value.*

*NOTE 3: An identifier may be the one or more attributes that determine if an identity passes or fails specific criteria.*

*EXAMPLE: A name of a club with a club membership number, a health insurance card number together with a name of the insurance company, an e-mail address, or a universally unique identifier (UUID) can all be used as identifiers. In a voter's register, the combination of attributes' name, address and date of birth is sufficient to unambiguously distinguish a voter.*

*Synonyms: unique identity, distinguishing identity. [ISO/IEC 24760-1]*

### Identity Authority

An entity that can make provable statements on the validity and/or correctness of one or more attribute values in an identity.

NOTE: An identity authority is typically associated with the domain, for instance, the domain of origin, in which the attributes, which the identity authority can make assertions on, have a particular significance.

**Identity Information**

*Set of values of attributes optionally with any associated metadata in an identity.*
*NOTE: In an information and communication technology system an identity is present as identity information. [ISO/IEC 24760-1]*

**Identity Management**

*Set of functions and capabilities (e.g., administration, management and maintenance, discovery, communication exchanges, correlation and binding, policy enforcement, authentication and assertions) used for:*
– *Assurance of identity information (e.g., identifiers, credentials, attributes),*
– *Assurance of the identity of an entity (e.g., users/subscribers, groups, user devices, organizations, network and service providers, network elements and objects, and virtual objects), and*
– *Enabling business and security applications.*
[ITU-T Y.2720]

**Identity Management System**

A system implementing identity management.

**Identity Provider**

A technical component that assures the unique identity and/or attributes of a principal according to the specified policy.

**IdP Discovery Service**

A service that allows the human user to discover the IdP during authentication.

**Certificate Authority**

A PKI certificate authority that is under control of the federation operator to certify public keys of federation entities.

### *Limited Linkability*

Technical inability stopping a service provider from sharing a principal's attributes with another SP to "link" service-specific data from multiple services, thus obtaining a wider view on a principal, which might violate privacy. But it is possible to link this data with the help of a third party, e.g., if the user consents.

### *Limited Observability*

Technical inability stopping an identity provider or attribute provider from collecting →connection data.

### *Metadata*

Configuration data required to automatically negotiate agreements between federation entities, like identifiers, binding support and endpoints, certificates and keys, cryptographic capabilities and security and privacy policies.

NOTE: The notion of metadata in the sense of data that reveals where and when a user accessed a service is described here as connection data.

### *Non-Disclosure*

Degree to which attributes are disclosed for a given transaction to a party other than the AP or SP.

### *Personally Identifiable Information*

*Any information (a) that can be used to identify the person to whom such information pertains, (b) from which such information can be derived, or (c) that is or might be directly or indirectly linked to a natural person.*

*NOTE: To determine whether a PII principal is identifiable, account should be taken of all the means which can reasonably be used by the entity holding the data, or by any other party, to identify that individual. [ISO/IEC 29100]*

### *Principal*

*Entity to which identity information pertains. [ITU-T X.811]*

NOTE 1: In particular, a principal interacts with an identity management system claiming an identity to a relying party in order to interact with services and access resources in a domain.

A principal is a real-world entity represented by a subject. The principal's identity information is bound to the issued credential.

NOTE 2: There are different classes of principals, such as individuals, persons in an organizational context, devices and software agents.

### *Pseudonymous Authentication*

An authentication is pseudonymous when the identifiers and other attributes of a principal released to an SP are not identifying and cannot be linked to the same principal's data in another SP. Pseudonymous identities may be re-identified with the help of a designated opener.

### *Reference Identifier (Reference ID)*

*Identifier in a domain that is intended to remain the same for the duration an entity is known in the domain and is not associated with another entity for a period specified in a policy after the entity ceases to be known in that domain.*

*NOTE 1: A reference identifier persists at least for the existence of the entity in a domain and may exist longer than the entity, e.g. for archival purposes.*

*NOTE 2: A reference identifier for an entity may change during the lifetime of an entity, at which point the old reference identifier is no longer applicable for that entity.*

*EXAMPLE: A driver license number that stays the same for an individual driver's driving life is a persistent identifier, which references additional identity information and that is a reference identifier. An IP address is not a reference identifier as it can be assigned to other entities. [ISO/IEC 24760-1]*

### *Relying Party*

*An entity that relies on an identity representation or claim by a requesting/asserting entity within some request context. [ITU-T Y.2720]*

### Service Broker

Federation entity that acts as a pseudonymizing proxy between identity authorities and SPs.

*Service Provider*

A system entity operated by the relying party that requires a requesting entity to be authenticated for a request to be processed.

NOTE: In this model the term SP can be replaced by SP affiliation.

*SP Affiliation*

A group of SPs allowed sharing a targeted ID for a principal. From a privacy perspective they act like a single SP, having a single attribute release policy, and usually having a single controller. Example: A single business service that is distributed to two SPs for technical reasons.

*Subject*

*The consumer of a digital service (a digital representation of a natural or juristic person, persona, group, organization, software service or device) described through claims [Cameron 2009].*

Subject is synonymous with claimant: *Entity that is or represents a principal for the purposes of authentication. NOTE: A claimant includes the functions necessary for engaging in authentication exchanges on behalf of a principal. [ITU-T X.1252, X.811]*

In the context of on-line gaming the concept is covered very well with the term avatar. In the context of data protection a data subject is any person or group of persons whose PII is being used (synonym to PII principal).

*Targeted Identifier (Targeted ID)*

A persistent, non-reassigned, privacy-preserving identifier for a principal shared between a pair of IdPs and SPs. An IdP uses the appropriate value of this attribute when communicating with a particular SP (or SP affiliation), and does not reveal that value to any other service provider except in limited circumstances. Many similar definitions can be found for EduPersonTargetedID[10].
Synonym: Persistent ID

*Temporary Identifier (Temporary ID)*

A targeted ID discarded after a session is terminated.

---

[10] e.g. SWITCH AAI attributes: http://www.switch.ch/it/aai/support/documents/attributes/